\documentclass[prd,nofootinbib,english]{revtex4}

\usepackage[utf8]{inputenc}
\usepackage{amsmath}
\usepackage{amsfonts,color}
\usepackage{amssymb,float}
\usepackage{color}
\usepackage{enumitem}
\usepackage{graphicx}
\usepackage{accents}
\usepackage{appendix}
\usepackage{hyperref}
\usepackage{url}
\usepackage{multirow}
\usepackage{orcidlink}

\hypersetup{
    colorlinks=true,
    linkcolor=blue,
    filecolor=magenta,      
    citecolor=red
}

\newcommand{\udt}[3]{#1^{#2}_{\phantom{#2}#3}}

\newcommand{\dut}[3]{#1_{#2}^{\phantom{#2}#3}}

\newcommand{\lc}[1]{\accentset{\circ}{#1}}

\begin{document}

\title{$f(T,B)$ cosmography for high redshifts}

\author{Celia Escamilla-Rivera\orcidlink{0000-0002-8929-250X}}
\email{celia.escamilla@nucleares.unam.mx}
\affiliation{Instituto de Ciencias Nucleares, Universidad Nacional Aut\'{o}noma de M\'{e}xico, 
Circuito Exterior C.U., A.P. 70-543, M\'exico D.F. 04510, M\'{e}xico.}

\author{Geovanny A. Rave-Franco\orcidlink{0000-0003-0638-8344}}
\email{	geovanny.rave@ciencias.unam.mx}
\affiliation{Instituto de Ciencias Nucleares, Universidad Nacional Aut\'{o}noma de M\'{e}xico, 
Circuito Exterior C.U., A.P. 70-543, M\'exico D.F. 04510, M\'{e}xico.}

\author{Jackson Levi Said\orcidlink{0000-0002-7835-4365}}
\email{jackson.said@um.edu.mt}
\affiliation{Institute of Space Sciences and Astronomy, University of Malta, Msida, Malta}
\affiliation{Department of Physics, University of Malta, Msida, Malta}


\begin{abstract}
In light of the statistical performance of cosmological observations, in this work we present the cosmography in $f(T,B)$ gravity. In this scenario we found a cosmological viable standard case that allows to reduce the degeneracy between several $f(T,B)$ models already proposed in the literature. Furthermore, we constrain this model using Pantheon SNeIa compilation, Cosmic Chronometers and a newly GRB calibrated data sample. We found that with an appropriate strategy for including the cosmographic parameter, we do produce a viable cosmology with our model within $f(T,B)$ gravity.
\end{abstract}

\maketitle


\section{Introduction}
\label{sec:intro}

General relativity (GR) has gone through decades of success as the fundamental theory of gravity of the concordance model of cosmology with observational consistency across all scales where measurements can be taken \cite{dodelson2003modern,Weinberg:1972kfs}. In cosmology, the combination of cold dark matter (CDM) together with a cosmological constant $\Lambda$ has led to the $\Lambda$CDM model of the Universe which is precisely compatible with observations related to the late-time accelerated expansion of the Universe \cite{Riess:1998cb,Perlmutter:1998np} as well as those from galactic dynamics \cite{alma991010731099705251}. Despite these accomplishments, $\Lambda$CDM is now facing a growing challenge from new observations related to the specific value of the Hubble constant \cite{Bernal:2016gxb,DiValentino:2020zio,DiValentino:2021izs}, as well as a noticeable tension in the growth of large scale structure \cite{Aghanim:2018eyx,DiValentino:2020vvd} over time. In addition, the theoretical problems within $\Lambda$CDM have been ever present during this period with the arbitrariness of the value of the cosmological constant being a serious issue in modern cosmology \cite{Weinberg:1988cp,Appleby:2018yci}, as well as the lack of direct measurements of dark matter in particle physics experiments despite enormous efforts \cite{Gaitskell:2004gd,Baudis:2016qwx,Bertone:2004pz}.

The growing observational and theoretical problems in $\Lambda$CDM have led to several efforts to modify the theory to better confront these developing challenges. While efforts to continue to allow for more exotic forms of matter continue \cite{Barboza:2008rh,Jassal:2005qc,Efstathiou:1999tm,Linder:2002et,Astier:2000as,Cooray:1999da,Escamilla-Rivera:2021boq,Colgain:2021pmf}, the possibility of modified gravity providing a possible resource to this problem has started to receive much more serious attention in the literature \cite{Clifton:2011jh,Capozziello:2011et,CANTATA:2021ktz}. There now exists an abundance of theories beyond GR in which potentially viable cosmological histories are possible. However, it is not enough for these gravitational models to produce fine-tuned models that agree with the numerical value of the late time accelerating Universe. In these generalized models, it is crucial that each component of the theory be connected with observational measurements. Cosmography \cite{Bamba:2012cp,Yang:2019vgk} opens that possibility where rather than repeatedly solving the Friedmann equations for slightly different assumptions and testing each of these choices, cosmographic parameters can be used since they are simply related to the derivatives of the scale factor \cite{Weinberg:1972kfs}. Thus, by just assuming a flat, homogeneous and isotropic cosmology, the distance--redshift relation can be used to produce successive derivative components of different generalized theories of gravity.

The weight of observational tensions in late time cosmology as well as the growing problems of dark matter particle physics and the foundational problems related to the structure of the theory has prompted several radical proposals to revisiting the origins of gravitation in order to meet these growing challenges. One of these possible descriptions of gravity is teleparallel gravity (TG) which embodies those theories in which the curvature associated with the Levi-Civita connection $\udt{\lc{\Gamma}}{\alpha}{\mu\nu}$ (over-circles represent quantities calculated with the Levi-Civita connection) is replaced with torsion \cite{Bahamonde:2021gfp,Aldrovandi:2013wha,Cai:2015emx,Krssak:2018ywd} through the teleparallel connection $\udt{\Gamma}{\alpha}{\mu\nu}$, which is curvature-less but continues to satisfy metricity. In this regime, all quantities that act as a measure of curvature will identically vanish such as the Riemann tensor $\udt{R}{\alpha}{\beta\mu\nu} \equiv 0$ \cite{Clifton:2011jh}. One consequential effect of this property is that the teleparallel Ricci scalar will be identically zero \cite{Aldrovandi:2013wha}, $R \equiv 0$. Thus, teleparallel theories must be based on a new geometric setting through the so-called torsion tensor \cite{Bahamonde:2021gfp} $\udt{T}{\alpha}{\mu\nu}$. An interesting feature of the torsion tensor is that by taking an appropriate series of contractions leads to a torsion scalar $T$ which produces a teleparallel equivalent of general relativity (TEGR) \cite{Aldrovandi:2013wha}. TEGR and GR differ by a boundary term $B$ in their Lagrangians but produce dynamically equivalent field equations.

The boundary term is the source of the largely fourth order nature of most modified theories of gravity that involve direct generalizations of the scalars based on the Riemann tensor, which is encapsulated in the Lovelock theorem \cite{Lovelock:1971yv} through its stringent conditions on the possible generalizations of GR that produce second order theories. Thus, in TG the class of theories which produce second order models immediately becomes drastically larger \cite{Gonzalez:2015sha,Bahamonde:2019shr} due to the decoupling of the boundary term in TEGR. Similar to GR, TEGR can be directly generalized through the many popular mechanisms by which modified theories of gravity can be formed. For instance, similar to $f(\lc{R})$ \cite{Sotiriou:2008rp,Faraoni:2008mf,Capozziello:2011et}, TEGR can be straightforwardly used to write a more arbitrary action based on an $f(T)$ gravity Lagrangian \cite{Ferraro:2006jd,Ferraro:2008ey,Bengochea:2008gz,Linder:2010py,Chen:2010va,Bahamonde:2019zea}. The decoupling of the Ricci scalar into the torsion scalar and boundary terms means that a more representative analogue to $f(\lc{R})$ gravity would be through $f(T,B)$ gravity \cite{Escamilla-Rivera:2019ulu,Bahamonde:2015zma,Capozziello:2018qcp,Bahamonde:2016grb,Paliathanasis:2017flf,Farrugia:2018gyz,Bahamonde:2016cul,Wright:2016ayu}. This limits to $f(\lc{R})$ gravity for the special limit in which $f(T,B) = f(-T+B) = f(\lc{R})$. $f(T,B)$ models have been explored in several regimes ranging from weak field solar system tests \cite{Farrugia:2020fcu,Capozziello:2019msc,Farrugia:2018gyz,Bahamonde:2020bbc}, to theoretical predictions from cosmology \cite{Bahamonde:2015zma,Bahamonde:2016grb,Bahamonde:2016cul,Bahamonde:2015zma}, as well as observational cosmology \cite{Escamilla-Rivera:2019ulu}. To this end, it is timely that further studies be performed to complement the work being done on determining which models of $f(T,B)$ gravity are more viable than others in terms of their confrontation with observations at all scales.

In this work, we explore the possibility of using cosmography as a tool to link observations on state parameters with the derivative term contributions to the $f(T,B)$ Lagrangian in the hope of further probing models which are observationally preferred. To this end, we first briefly introduce $f(T,B)$ gravity in Sec.~\ref{sec:fTB} where we describe the nuances of the theory. In Sec.~\ref{sec:cosmo}, we show how the different components of the $f(T,B)$ model can connect to the different cosmographic parameters at play. Through this analysis, we explore the extent to which this analysis can be used to formulate a cosmologically viable model in this scheme. Finally, we close with a brief summary of our work in Sec.~\ref{sec:conclusions}. We also have included two appendices to better explain the details of this work. In Appendix \ref{app:calculations} we discuss the theoretical background of the model analysis we explore in this work, while in Appendix \ref{app_pheno} we explore some technical details related to how phenomenology.


\section{\texorpdfstring{$f(T,B)$}{} gravity on basis}
\label{sec:fTB}

GR is based on the curvature associated with the Levi-Civita connection $\udt{\lc{\Gamma}}{\sigma}{\mu\nu}$ (recalling that over-circles refer to any quantities based on the Levi-Civita connection) while TG is built on its exchange with the teleparallel connection $\udt{\Gamma}{\sigma}{\mu\nu}$ \cite{Aldrovandi:2013wha,Bahamonde:2021gfp,Cai:2015emx,Krssak:2018ywd}. This alternative approach to building the foundations of gravitation is arrived at by first replacing the metric tensor $g_{\mu\nu}$ as the fundamental dynamical variable with the tetrad $\udt{e}{A}{\mu}$ and spin connection $\udt{\omega}{B}{C\nu}$. In all instances, Greek indices refer to coordinates on the general manifold while Latin ones refer to coordinates on the local Minkowski space. Thus, tetrads connects both spaces and can raise and lower indices between the different spaces \cite{RevModPhys.48.393}
\begin{align}
    g_{\mu\nu} = \udt{e}{A}{\mu}\udt{e}{B}{\nu} \eta_{AB}\,,& &\text{and}& &\eta_{AB} = \dut{E}{A}{\mu}\dut{E}{B}{\nu} g_{\mu\nu}\,,\label{metric_tetrad_eq}
\end{align}
where orthogonality conditions produce the conditions
\begin{align}
    \udt{e}{A}{\mu}\dut{E}{B}{\mu} = \delta_B^A\,,& &\text{and}& &\udt{e}{A}{\mu}\dut{E}{A}{\nu} = \delta_{\mu}^{\nu}\,.
\end{align}
However, there exists an infinite number of solutions of this equation which occurs due to the freedom in choosing the local Lorentz frame. This freedom is embodied in the TG flat spin connection.

The tetrad and flat spin connection pair can then be used to define the teleparallel connection as \cite{Cai:2015emx,Krssak:2018ywd}
\begin{equation}
    \Gamma^{\lambda}{}_{\nu\mu}=\dut{E}{A}{\lambda}\partial_{\mu}\udt{e}{A}{\nu}+\dut{E}{A}{\lambda}\udt{\omega}{A}{B\mu}\udt{e}{B}{\nu}\,,
\end{equation}
which is curvature-less and satisfies metricity \cite{Hohmann:2021fpr}. For the gauge choice where the spin connection vanishes, this is called the Weitzenb\"{o}ck gauge \cite{Weitzenbock1923}. Thus, the tetrad spin connection pair represent the fundamental variables of the theory and so produce independent field equations.

Analogous to GR, TG is build on the concept of tensor objects defined on a Riemann manifold. However, the Riemann tensor $\udt{\lc{R}}{\alpha}{\beta\mu\nu}$ is a measure of curvature, and so if the Levi-Civita connection is replaced with the teleparallel connection then its components will organically vanish $\udt{R}{\alpha}{\beta\mu\nu} \equiv 0$. Thus, a torsion tensor is defined to give a measure of torsion \cite{Aldrovandi:2013wha,ortin2004gravity}
\begin{equation}
    \udt{T}{A}{\mu\nu} := 2\udt{\Gamma}{A}{[\nu\mu]}\,,
\end{equation}
where the brackets denote the antisymmetric operator, and this acts as a measure of the field strength of gravity in TG \cite{Bahamonde:2021gfp,Aldrovandi:2013wha}. Taking a suitably chosen addition of contractions leads to the torsion scalar \cite{Krssak:2018ywd,Cai:2015emx,Aldrovandi:2013wha,Bahamonde:2021gfp}
\begin{equation}
    T:=\frac{1}{4}\udt{T}{\alpha}{\mu\nu}\dut{T}{\alpha}{\mu\nu} + \frac{1}{2}\udt{T}{\alpha}{\mu\nu}\udt{T}{\nu\mu}{\alpha} - \udt{T}{\alpha}{\mu\alpha}\udt{T}{\beta\mu}{\beta}\,,
\end{equation}
which is arrived at by demanding that $T$ is equivalent to the Ricci scalar (up to a total divergence term). Analogous to the Ricci scalar being dependent only on the Levi-Civita connection, the torsion scalar is only dependent on the teleparallel connection. The teleparallel Ricci scalar will identically vanish due to it being torsion-less. However, this relation does show the equivalence of the regular Ricci scalar and the torsion scalar through \cite{Bahamonde:2015zma}
\begin{equation}\label{LC_TG_conn}
    R=\lc{R} + T - B = 0\,,
\end{equation}
where $B$ is a boundary term defined by
\begin{equation}
    B:=2\mathring{\nabla}_{\mu}\left(T^{\mu}\right)\,,
\end{equation}
and where $e = \text{det}\left(\udt{e}{A}{\mu}\right)=\sqrt{-g}$ is the tetrad determinant. The torsion scalar is the Lagrangian of TEGR and through this relation, we have a guarantee that the equations of GR and TEGR will be dynamically equivalent to each other.

In regular curvature-based gravity, one of the most popular approaches to modifying GR is that of $f(\lc{R})$ gravity \cite{Sotiriou:2008rp,Capozziello:2011et} in which the Einstein-Hilbert action Ricci scalar is generalized to an arbitrary function thereof. The decoupling of the second and fourth order contributions to the Ricci scalar means that the analogous setup would be contained in a broader $f(T,B)$ generalization \cite{Bahamonde:2015zma,Capozziello:2018qcp,Bahamonde:2016grb,Paliathanasis:2017flf,Farrugia:2018gyz,Bahamonde:2016cul,Bahamonde:2016cul,Wright:2016ayu,Zubair:2021gve,Zubair:2020wyu,Zubair:2018wyy}. The action for this theory can then be written as
\begin{equation}\label{f_T_action}
    \mathcal{S}_{f(T,B)} = \frac{1}{2\kappa^2}\int d^4 x\, ef(T,B) + \int d^4 x\, e\mathcal{L}_m\,.
\end{equation}
where the $f(\lc{R})$ limits occurs for $f(T,B)=f(-T+B)=f(\lc{R})$ and TEGR is found when $f(T,B)=-T$.

For a flat Friedmann--Lema\^{i}tre--Robertson--Walker (FLRW) metric in $f(T,B)$ gravity we can take the tetrad
\begin{equation}\label{flrw_tetrad}
    \udt{e}{A}{\mu}=\textrm{diag}(1,a(t),a(t),a(t))\,,
\end{equation}
which is in the Weitzenb\"{o}ck gauge \cite{Krssak:2015oua,Tamanini:2012hg}, and where $a(t)$ is the scale factor. This tetrad reproduces the FLRW metric to give
\begin{equation}
    ds^2=dt^2-a(t)^2(dx^2+dy^2+dz^2)\,,
\end{equation}
through Eq.~(\ref{metric_tetrad_eq}). The corresponding torsion scalar turns out to be
\begin{equation}
    T = 6H^2\,,
\end{equation}
while the boundary term is given by 
\begin{equation}
    B =6(3H^2+\dot{H})\,,
\end{equation}
which together produce the regular Ricci scalar, namely $\lc{R}=-T+B = 6(\dot{H} + 2H^2)$. The analogous Friedmann equations for this tetrad are given by
\begin{eqnarray}
-3H^2(3f_B + 2f_T) + 3H\dot{f}_B - 3\dot{H}f_B + \frac{1}{2}f &= \kappa \rho, \label{eq:friedmann_mod} \\
-\left(3H^2 + \dot{H} \right)(3f_B + 2f_T) - 2H\dot{f}_T + \ddot{f}_B + \frac{1}{2}f &= -\kappa P.\label{eq:friedmann_mod2}
\end{eqnarray}
where overdots represent time derivatives, while $\rho$ and $P$ are respectively the energy density and pressure of a perfect fluid, and where underscore denotes partial derivatives (such as $f_T = \partial f/\partial T$ and $f_B = \partial f/\partial B$).


\section{Extended cosmography for \texorpdfstring{$f(T,B)$}{} gravity}
\label{sec:cosmo}

Cosmography offers a way to parameterize the evolution of the Universe through its state parameters which can be used to infer progressively higher order derivatives of our gravitational Lagrangian, namely $f(T,B)$. This provides a directly observationally-driven approach to attempting to resolve the open questions surrounding possible modifications to the Einstein-Hilbert action. To this end, we employ the standard setting of defining the cosmographic parameters up to the seventh order in terms of derivatives of the scalar factor, namely \cite{Capozziello:2011hj,Aviles:2012ay}
\begin{eqnarray}
H = \frac{1}{a}\frac{da}{dt}\,, \label{eq:Hparameter}
\quad
q = -\frac{1}{a}\frac{d^2a}{dt^2}H^{-2}\,, \label{eq:qparameter}
\quad
j = \frac{1}{a}\frac{d^3a}{dt^3}H^{-3}\,,\label{eq:jparameter} \\
s = \frac{1}{a}\frac{d^4a}{dt^4}H^{-4}\,,\label{eq:sparameter} 
\quad
l = \frac{1}{a}\frac{d^5a}{dt^5}H^{-5}\,,\label{eq:lparameter} 
\quad 
\tilde{m} = \frac{1}{a}\frac{d^6a}{dt^6}H^{-6}\,,\label{eq:mparameter} 
\quad
n = \frac{1}{a}\frac{d^7a}{dt^7}H^{-7}\,, \label{eq:nparameter}
\end{eqnarray}
which all describe the kinematic evolution of the Universe. By considering the Friedmann equations of $f(T,B)$ \cite{Franco:2020lxx,Escamilla-Rivera:2019ulu,Caruana:2020szx} we can relate the derivatives of the Lagrangian with the evolution of the Universe. On a practical level, this occurs due to the relations between the contributing torsion scalar and boundary term, and the derivatives of the Hubble parameter. We thus write down these calculations for clarity for the work that follows
\begin{eqnarray}
&\dot{T} = 12H\dot{H}\,,  &\dot{B} = 3\dot{T} + 6\ddot{H}\,,\\ \label{eq:FirstDotTB}
&\ddot{T} = 12 \left[\dot{H}^2 + H\ddot{H}\right]\,, &\ddot{B}=3\ddot{T} + 6\dddot{H}\,, \\ \label{eq:SecondDotTB}
&\dddot{T} = 12\left[3\dot{H}\ddot{H} + H\ddot{H}\right]\,, &\dddot{B} = 3\dddot{T} + 6H^{\text{(iv)}}\,,\\\label{eq:ThirdDotTB}
&T^{\text{(iv)}} = 12\left[3\ddot{H}^2 + 4\dot{H}\dddot{H} + HH^{\text{(iv)}} \right]\,, &B^{\text{(iv)}}=3T^{\text{(iv)}} + 6H^{\text{(v)}}\,,\label{eq:FourthDotTB} \\
&T^{\text{(v)}} = 12[10\ddot{H}\dddot{H} + 5\dot{H}H^{\text{(iv)}} + HH^{\text{(v)}}]\,,  &B^{\text{(v)}}=3T^{\text{(v)}} + 6H^{\text{(vi)}}\,,\label{eq:FifthDotTB}
\end{eqnarray}
which can then be written in terms of the cosmographic parameters \eqref{eq:Hparameter}-\eqref{eq:nparameter} as
\begin{equation}\label{eq:TyB}
\begin{aligned}[c]
 T &= 6H^2\,, \\
\dot{T} &= -12 H^3 \left(q+1\right)\,, \\
\ddot{T} &= 12 H^4 \left[j+q \left(q+5\right)+3\right]\,, \\
\dddot{T} &= 12 H^5 \left[-j \left(3 q+7\right) \right. \\ & \text{\hspace{0.6cm} } \left.-3 q \left(4 q+9\right)+s-12\right]\,, \\
T^{\text{(iv)}} &= 12 H^6 \left[j \left(3 j+44 q+48\right)\right. \\ & \text{\hspace{0.6cm} }\left.+l-\left(4 q+9\right) s \right. \\ & \text{\hspace{0.6cm} }\left.+3 q \left(4 q^2+39 q+56\right)+60\right]\,, \\
T^{\text{(v)}} &= -12 H^7 \left(50 j^2+10 j (q (8 q+51)-s+36)\right.\\&\left.+l (5 q+11)-\tilde{m}-5 (14 q+15) s\right. \\ &\left.+30 q (q+2) (9 q+20)+360\right)\,,
\end{aligned}
\quad
\begin{aligned}[c]
B &=-6 H^2 \left(q-2\right)\,,\\
\dot{B} &= 6 H^3 \left(j-3 q-4\right)\,,\\
\ddot{B}&=6 H^4 \left[2 j+3 q, \left(q+6\right)+s+12\right]\,, \\
\dddot{B} &= 6 H^5 \left[-2 j \left(4 q+11\right)+l \right.\\ & \text{\hspace{0.4cm} }\left.-6 \left(q \left[7 q+17\right)+8\right]+s\right]\,,\\
B^{\text{(iv)}}&=6 H^6 \left[24 j \left(6 q+7\right)+8 j^2\right. \\ &  \text{\hspace{0.4cm} }\left.+\tilde{m}-3 \left(3 q+8\right) s\right. \\ & \text{\hspace{0.4cm} }\left.+6 q \left(q \left[7 q+72\right]+108\right)+240\right]\,, \\
B^{\text{(v)}}&=-6 H^7 (160 j^2 + \tilde{m} - n + 3 l (8 + 3 q)\\ &+ 
   30 (q (156 + 3 q (48 + 11 q) - 7 s) - 8 (-6 + s))\\ &+ 
   5 j (264 + 18 q (20 + 3 q) - 5 s))\,.
\end{aligned}
\end{equation}
Here, we have the gravitational scalars directly related with the cosmographic parameters which makes it possible to infer their values through observational data. At this point, we consider a simple model $f(T,B)=-T + \tilde{f}(T,B)$ as a power law model $\tilde{f}(T,B)=\tilde{t}_0T^m + \tilde{b}_0B^k$. Power law models are interesting ways to probe new physics in these settings, but this model also exhibits the property that any cross derivatives will vanish. the remainder of the contributing derivatives are shown in Appendix \ref{app:calculations}. In $f(T,B)$ gravity, the power law model is motivated not only for its ease of purpose in the cosmographic setting but also for its observational and theoretical advantages. In Ref.~\cite{Escamilla-Rivera:2019ulu}, this model is shown to model cosmic chronometer, Pantheon and baryonic acoustic oscillation data in the late Universe. Moreover, in Ref.~\cite{Caruana:2020szx} this model seems to be a genetic solution in many instances some early Universe physics, while the power law model also seems to have some interesting features for an astrophysics setting \cite{Bahamonde:2021srr,Farrugia:2020fcu}.

The Friedmann equations \eqref{eq:friedmann_mod} and \eqref{eq:friedmann_mod2} can be now be rewritten as
\begin{eqnarray}
3H^2 &=& \kappa^2 \rho +  3H^2\left(3f_B + 2f_T\right) - 3H\dot{f}_B + 3\dot{H}f_B - \frac{1}{2}f\label{friedmann1} \,,\\
2 \dot{H} &=& -\kappa^2\rho(1+\omega) + 2\dot{H}f_T + 2H\dot{f}_T + 3H\dot{f}_B - \ddot{f}_B \label{friedmann2} \,,
\end{eqnarray}
where $\omega = p/\rho$ is the equation of state for matter. If we compute higher order derivatives of $H$ in a dust model $\omega=0$, we obtain
\begin{eqnarray}
2\ddot{H}&=&3H\kappa^2\rho + 2\ddot{H}f_T + 2H\ddot{f}_T + 4\dot{H}\dot{f}_T + 3\dot{H}\dot{f}_B + 3H\ddot{f}_B - \dddot{f}_B\,,\label{friedmann3}  \\
2\dddot{H}&=&3\dot{H}\kappa^2\rho - 9H^2\kappa^2\rho + 2\dddot{H}f_T + 2H\dddot{f}_T + 6\ddot{H}\dot{f}_T + 6\dot{H}\ddot{f}_T + 3\ddot{H}\dot{f}_B\label{friedmann4}  \\&&+ 6\dot{H}\ddot{f}_B + 3H\dddot{f}_B - f^{\text{(iv)}}_B, \notag \\
2H^{\text{(iv)}} &=& 3\ddot{H}\kappa^2\rho - 27H\dot{H}\kappa^2\rho + 27H^3\kappa^2\rho + 2H^{\text{(iv)}}f^{\text{(iv)}}_T + 8\dddot{H}\dot{f}_T + 8\dot{H}\ddot{f}_T + 12\ddot{H}\ddot{f}_T\\ \notag && + 9\ddot{H}\ddot{f}_B +3\dddot{H}\dot{f}_B + 9\dot{H}\dddot{f}_B + 3Hf^{\text{(iv)}}_B - f^{\text{(v)}}_B\,.
\end{eqnarray}

These expressions allow us to write the cosmographic parameters in terms of the arbitrary $f(T,B)$ Lagrangian and its derivatives, explicitly, they can be written as 
\begin{eqnarray}
q=-1+\frac{1}{3 f_B+2 f_T-2}\left[-\frac{\ddot{f}_B}{H^2}+9 f_B-\frac{f}{2 H^2}+\frac{2 \dot{f}_T}{H}+6 f_T-3\right], \label{eq:q_interms} \\
\Omega_{\text{m}} = 1-3f_B - 2f_T + \frac{1}{H}\dot{f}_B + (1+q)f_B + \frac{1}{6H^2}f\,, \\
j = -2-3q + \frac{1}{2(1-f_T)}\Bigg[ -\frac{\dddot{f}_B}{H^3}+\frac{3 \ddot{f}_B+2 \ddot{f}_T}{H^2}-\frac{(q+1) \left(3 \dot{f}_B+4 \dot{f}_T\right)}{H}+9 \Omega_{\text{m}} \Bigg]\,,\\
s = 4j + 3q(q+4) + 6 +  \frac{1}{2(1-f_T)}\Bigg[-\frac{f^{\text{(iv)}}_B}{H^4}+\frac{3 \dddot{f}_B+2 \dddot{f}_T}{H^3}-\frac{6 (q+1) \left(\ddot{f_B}+\ddot{f_T}\right)}{H^2} \nonumber \\  +\frac{(j+3
   q+2) \left(3 \dot{f}_B+6 \dot{f}_T\right)}{H}-(9 (q+1)+27) \Omega_{\text{m}} \Bigg]\,, \\
l = 5s - 10(q+2)j - 30(q+2)q - 24 + \frac{1}{2(1-f_T)}\Bigg[-\frac{f^{\text{(v)}}_B}{H^5}+\frac{3 f^{\text{(iv)}}_B+2 f^{\text{(iv)}}_T}{H^4} 
\nonumber \\ \notag -\frac{(q+1) \left(9 \dddot{f}_B+8
   \dddot{f}_T\right)}{H^3}+\frac{(j+3 q+2) \left(9 \ddot{f}_B+12 \ddot{f}_T\right)}{H^2}+\frac{\left(3 \dot{f}_B+8 \dot{f}_T\right) (-4 j-3 q
   (q+4)+s-6)}{H}\\ +\Omega_{\text{m}}  (9 (j+3 q+2)+81 (q+1)+81)\Bigg]\,,\label{eq:l_interms}
\end{eqnarray}
where $\Omega_{\rm m}$ is the density parameter for matter.

However, when substituting the derivatives given by the Eqs.\eqref{eq:f_T}--\eqref{eq:f_biv} into the above expression, we obtain a system of nonlinear relations that makes it impossible to isolate the cosmographic parameters in a general model for arbitrary $k$ and $m$. To highlight this issue, we can see that the deceleration parameter when substituting the Eqs.\eqref{eq:f_T}--\eqref{eq:f_biv}, becomes
{\footnotesize
\begin{align} 
q &=-1 + 3 \left(\frac{\tilde{b}_0 6^{k-1} (k-1) k \left(-H^2 (q-2)\right)^k \left(j^2 (k-2)-2 j k (3 q+4)+10 j (q+2)+k (3 q+4)^2-3 q^3-30 q^2-q s-24 q+2 s-8\right)}{H^2 (q-2)^3} \right. \label{eq:q implicit}\\ \notag &\left.-\frac{\tilde{b}_0
   6^k \left(-H^2 (q-2)\right)^k+6^m \tilde{t}_0 \left(H^2\right)^m}{2 H^2}+9 \tilde{b}_0 k \left(18 H^2-6 H^2 (q+1)\right)^{k-1}-2^{m+1} 3^{m-1} (m-1) m (q+1) \tilde{t}_0 \left(H^2\right)^{m-1}\right. \\ \notag & \left. +6^m m
   \tilde{t}_0 \left(H^2\right)^{m-1}-3\right) \div \Bigg(9 \tilde{b}_0 k \left(18 H^2-6 H^2 (q+1)\right)^{k-1}+6^m m \tilde{t}_0 \left(H^2\right)^{m-1}-6\Bigg), 
\end{align}
}
which can not be isolated for arbitrary values of $k$ and $m$. The same issue holds for the other cosmographic parameters. Thus, the cosmographic parameters are inherently interrelated in this sense.

\subsection{Standard Case}
\label{sec:case}

As an example of a standard case, it is possible to isolate the cosmographic parameters in the simplest case with $m=1$ and $k=2$. In this context, the equation \eqref{eq:q implicit} simplifies to
\begin{align}
q = \frac{6 \tilde{b}_0 H^2 (4 j+9 q (q+4)+2 (s+6))-\tilde{t}_0+1}{36 \tilde{b}_0 H^2 (q-2)-2 \tilde{t}_0+2}\,,\label{eq:dece_par_sc}
\end{align}
where it is possible to solve for $q$ as
\begin{eqnarray}
\label{eq:qfixed}
q_1 &= \frac{\left(\tilde{t}_0-1\right) \left(\sqrt{1-\frac{18 \tilde{b}_0 H^2 \left(12 \tilde{b}_0 H^2 (2 j+s-90)-17 \left(\tilde{t}_0-1\right)\right)}{\left(\tilde{t}_0-1\right){}^2}}-1\right)}{18 \tilde{b}_0 H^2}-8\,, \\
q_2 &= -\frac{\left(\tilde{t}_0-1\right) \left(\sqrt{1-\frac{18 \tilde{b}_0 H^2 \left(12 \tilde{b}_0 H^2 (2 j+s-90)-17 \left(\tilde{t}_0-1\right)\right)}{\left(\tilde{t}_0-1\right){}^2}}+1\right)}{18 \tilde{b}_0 H^2}-8\,.
\end{eqnarray}
Under the same assumption, is is also possible to isolate the other cosmographic parameters which are given by
{\small
\begin{align}\label{eq:cosmo-fixed}
\Omega_{\text{m}} &= \frac{2 \tilde{b}_0 \left(6 H^3 (j+3 q+2)-36 H^3 (q+1)\right)}{H}+\frac{\tilde{b}_0 \left(18 H^2-6 H^2 (q+1)\right)^2+6 H^2 \tilde{t}_0}{6 H^2} \nonumber \\ &+2 \tilde{b}_0 (q+1) \left(18 H^2-6 H^2 (q+1)\right)-6 \tilde{b}_0
   \left(18 H^2-6 H^2 (q+1)\right)-2 \tilde{t}_0+1 \\[10pt]
j &= \frac{-\frac{6 \tilde{b}_0 H^2 l}{\tilde{t}_0-1}+\frac{333 \tilde{b}_0 H^2 q^2}{\tilde{t}_0-1}+\frac{1008 \tilde{b}_0 H^2 q}{\tilde{t}_0-1}+\frac{12 \tilde{b}_0 H^2 s}{\tilde{t}_0-1}+\frac{252 \tilde{b}_0 H^2}{\tilde{t}_0-1}+3 q-\frac{5}{2}}{-\frac{6 \tilde{b}_0
   H^2 (5 q+34)}{\tilde{t}_0-1}-1} \\[10pt]
   s &= \Bigg(-\frac{30 \tilde{b}_0 H^2 j^2}{\tilde{t}_0-1}-\frac{54 \tilde{b}_0 H^2 j (21 q+32)}{\tilde{t}_0-1}+\frac{18 \tilde{b}_0 H^2 l}{\tilde{t}_0-1}-\frac{6 \tilde{b}_0 H^2 m}{\tilde{t}_0-1}-\frac{333 \tilde{b}_0 H^2 q^3}{\tilde{t}_0-1}-\frac{4104
   \tilde{b}_0 H^2 q^2}{\tilde{t}_0-1} \\ \notag &-\frac{6588 \tilde{b}_0 H^2 q}{\tilde{t}_0-1}-\frac{1584 \tilde{b}_0 H^2}{\tilde{t}_0-1}-4 j-\frac{3}{2} (2 q+5) q+12\Bigg)\div\Bigg(-\frac{18 \tilde{b}_0 H^2 (q+7)}{\tilde{t}_0-1}-1\Bigg) \\[10pt]  
   l &= \Bigg( \frac{1194 \tilde{b}_0 H^2 j^2}{\tilde{t}_0-1}+\frac{3 \tilde{b}_0 H^2 j (9 q (79 q+626)-26 s+4644)}{\tilde{t}_0-1}+\frac{60 \tilde{b}_0 H^2 m}{\tilde{t}_0-1}-\frac{6 \tilde{b}_0 H^2 n}{\tilde{t}_0-1}\\\notag&+\frac{18 \tilde{b}_0 H^2 (-76 q s+6 q
   (q (86 q+411)+444)-105 s+598)}{\tilde{t}_0-1}+\frac{1}{2} j (20 q+31)+6 q (5 q+1)-5 s-66 \Bigg)\\ \notag & \div \Bigg( -\frac{90 \tilde{b}_0 H^2}{\tilde{t}_0-1}-1 \Bigg)
\end{align}
}

In a general FLRW spacetime this can be described for small distances by the luminosity distance which is expressed as
\begin{eqnarray}
d_L(z) &=& \frac{c}{H_0}[z+\frac{1}{2}(1-q_0)z^2-\frac{1}{6}(1-q_0-3q_0^2+j_0+ \frac{kc^2}{H_0^2a^2(t_0)})z^3+ \nonumber \\ && +\frac{1}{24}[2-2q_0-15q_0^2-15q_0^3+5j_0+10q_0j_0+
s_0+\frac{2kc^2(1+3q_0)}{H_0^2a^2(t_0)}]z^4+\ldots]\,,
\end{eqnarray}
where the cosmographic parameters are defined as
\begin{eqnarray}
H_0 &\equiv&\frac{1}{a(t)} \frac{da(t)}{dt}|_{t=t_0} \equiv \frac{\dot{a}(t)}{a(t)}|_{t=t_0}~,\label{eq:H0}\\
q_0&\equiv&-\frac{1}{H^2}\frac{1}{a(t)}\frac{d^2a(t)}{dt^2}|_{t=t_0}\equiv-\frac{1}{H^2}\frac{\ddot{a}(t)}{a(t)}|_{t=t_0}~,\label{eq:q0}\\
j_0&\equiv& \frac{1}{H^3}\frac{1}{a(t)}\frac{d^3a(t)}{dt^3}|_{t=t_0}\equiv \frac{1}{H^3}\frac{a^{(3)}(t)}{a(t)}|_{t=t_0}~,\label{eq:j0}\\
s_0&\equiv& \frac{1}{H^4}\frac{1}{a(t)}\frac{d^4a(t)}{dt^4}|_{t=t_0}\equiv \frac{1}{H^4}\frac{a^{(4)}(t)}{a(t)}|_{t=t_0}~.\label{eq:s0}
\end{eqnarray}

Notice that depending on the degree of Taylor series considered, the cosmographic expressions derived could produce degenerate results. Moreover, this may lead to a misunderstanding in the definition of the proper distance. On the other hand, we can consider observations from recent surveys to constrain the impact of this degeneracy through truncation of the Taylor series. In thi way, we can use the luminosity distance relation to relate cosmographic parameters with observational data. In this context, we consider the luminosity distance as the most direct choice in the measurements of distance for Type Ia Supernovae (SNeIa) and gamma-ray bursts (GRBs), along the Hubble flow using cosmic chronometers (CCs).

To fit the luminosity distance redshift relation using this data and to solve the convergence issue at high-$z$ via an adequate truncation of the series, it is useful to rewrite $d_L$ as a function of the proposed variable $y=z/(1+z)$. In this way, we can map $z\in(0,\infty)$ into $y\in(0,1)$, and retrieve an optimal
behaviour for the Taylor series at any redshift. This new variable $y$ does not change the definition of the cosmographic parameters and the luminosity distance at fourth order in such variable can be written as
\begin{eqnarray}
d_L(y)=&&\frac{c}{H_0}\left\{y-\frac{1}{2}(q_0-3)y^2+\frac{1}{6}\left[12-5q_0+3q^2_0-(j_0+\Omega_0)\right]y^3+\frac{1}{24}\left[60-7j_0-\right.\right.\nonumber
  \\ &&\left.\left.-10\Omega_0-32q_0+10q_0j_0+6q_0\Omega_0+21q^2_0-15q^3_0+s_0\right]y^4+\mathcal{O}(y^5)\right\}\,,
  \label{eq:distance}
\end{eqnarray}
where $\Omega_0=1+kc^2/H_0^2a^2(t_0)$, is the total energy density at current times. Therefore, we can write the luminosity distance logarithmic Hubble relation as
\begin{eqnarray}
    \ln{\left[\frac{d_L(y)}{y~}\right]} \rm Mpc^{-1} =&&\frac{\ln{10}}{5}\left[\mu(y)-25\right]-\ln{y}=\ln{\left[\frac{c}{H_0}\right]}
    -\frac{1}{2}(q_0-3)y+\frac{1}{24}\left[21-4(j_0+\Omega_0)+\right.\nonumber\\ 
    &&\left.+q_0(9q_0-2)\right]y^2+\frac{1}{24}\left[15+4\Omega_0(q_0-1)+j_0(8q_0-1) -5q_0+2q^2_0  \right.\nonumber\\ 
    &&\left.  -10q^3_0+s_0\right]y^3+\mathcal{O}(y^4)\,,\label{eq:y-exp}
\end{eqnarray}
and the corresponding equation for the distance modulus is 
\begin{eqnarray}
    \mu(y)=&&25+\frac{5}{\ln{10}}\left\{\ln{\left[\frac{c}{H_0}\right]}+\ln{y} - \frac{1}{2}(q_0-3)y+\frac{1}{24}\left[21-4(j_0+\Omega_0)+q_0(9q_0-2)\right]y^2+\right.\nonumber\\ 
    &&\hspace{-.5cm}+\left.\frac{1}{24}\left[15+4\Omega_0(q_0-1)+j_0(8q_0-1)-5q_0+2q^2_0-10q^3_0+s_0\right]y^3+\mathcal{O}(y^4)\right\}\,,\label{eq:modulus}
\end{eqnarray}
which is important for SNeIa data. Notice that the higher order the term in the expansion of $d_L$, the better the fit to the data sample we can obtain since there will be more free parameters -- but of course this comes at the cost of greater degeneracy in these parameters. However, for a given data there will be an upper bound on the order of the series which gives a statistically significant fit to those data.

\section{Observational cosmographic constraints}
\label{sec:data}

To perform the statistical analyses of the cosmography found and understand current constraints, we need to focus on specific data sets and likelihood functions.
In this analysis we are going to consider three samples: a SNeIa compilation, a newly calibrated GRB sample and the CCs to constrain the cosmographic parameters from the standard $f(T,B)$ model.

\begin{itemize}
    \item \textbf{Pantheon compilation.} The recent SNeIa compilation known as the Pantheon sample (SN) \cite{Scolnic:2017caz} consists of 1048 SNeIa compressed in 40 redshift bins. As already analysed in the literature, SNeIa can provide estimates of the distance modulus, $\mu$, the theoretically predicted value of which is related to the luminosity distance $d_L$ we obtain in Eq.(\ref{eq:distance}) and Eq.(\ref{eq:modulus}) as follows
\begin{equation}\label{eq:lum}
\mu(z)= 5\log{\left[\frac{d_L (z)}{1 \text{Mpc}}\right]} +25,
\end{equation}
where the luminosity distance $d_L$ is given in Mpc. In this distance modulus expression we should include the nuisance parameter, $\mathcal{M}$, as an unknown offset of the supernovae absolute magnitude (and including other possible systematics), which can also be degenerate with the value of $H_0$. As is standard, we assume spatial flatness and suppose that $d_L$ can be related to the comoving distance $D$ using
$d_{L} (z) =\frac{c}{H_0} (1+z)D(z),$
where $c$ is the speed of light. We obtain
$D(z) =\frac{H_0}{c}(1+z)^{-1}10^{\frac{\mu(z)}{5}-5}.$
The normalised Hubble function $H(z)/H_0$ is derived from the inverse of the derivative of $D(z)$ with respect to $z$ so that we can write down
\begin{equation}
D(z)=\int^{z}_{0} \frac{H_0 d\tilde{z}}{H(\tilde{z})}, \label{eq:dist}
\end{equation}
where $H_0$ is the value of the Hubble constant which we consider as a prior to normalise $D(z)$. For our sample, we calibrated the data by using a value obtained from late universe measurements corresponding to $H_0 =73.8 \pm 1.1 \, \, \mbox{km s}^{-1} \mbox{Mpc}^{-1}, $ from SH0ES + H0LiCOW, with the corresponding value for the nuisance parameter $M=-19.24 \pm 0.07$.
    \item \textbf{New GRB dataset.} After the reconstruction/prior calibration of $d_L$ from SN, we can use them to calibrate the luminosity correlations of the GRB data set. The correlations obtained from this calibration can be expressed by considering a generic exponential form $R = AQ^b$, where this assumption is derived by consider the X-ray light curves of GRBs constructed from the combination of Burst Alert Telescope and X- ray telescope data in the way described in \cite{2010GRBSHi} and fitted using one or two components from the optional afterglow via the parametrised Amati relation and the  Ghirlanda relation \cite{Wang:2019fie}, where this expression can be re-expressed in linear form as $y= a+ bx$, with $y\equiv \log R$, $x\equiv \log Q$ and $a= \log A$. Also, the six luminosity correlations measured are reported in \cite{Escamilla-Rivera:2021vyw}.

To calibrate these six expressions with a SNeIa sample (in particular, for Pantheon sample), we consider that GRBs radiate isotropically by computing their bolometric peak flux, where uncertainty on $L$ propagates from the uncertainties on bolometric peak flux $P_{\text{bolo}}=L/4\pi d^{2}_{L}$,
and $d_L$ the luminosity distance of the supernovae. With this new sample we calibrated the luminosity correlations by maximizing the likelihood \cite{DAgostini:2005mth}
\begin{equation}
\mathcal{L}(\sigma_{\text{int}},a,b) \propto \prod_{i} \frac{1}{\sqrt{\sigma^{2}_{\text{int}}+\sigma^2_{yi} +b^2 \sigma_{xi}}} \times
\text{exp} \left[ -\frac{(y_i -a - bx_i)^2}{2(\sigma^2_{\text{int}} +\sigma^2_{yi} +b^2 \sigma^2_{xi})} \right].
\end{equation}
The best-fitting parameters and their uncertainties,$(a , b, \sigma_{\text{int}})$, are reported in Tables 1 and 2 from \cite{Escamilla-Rivera:2021vyw}.

\item Cosmic Chronometers. We consider a sample of 31 model-independent measurements which use the differential age method proposed by \cite{Jimenez:2001gg}.

\end{itemize}

In order to determine viable cosmographic constraints that can reproduce the observed cosmic acceleration for the standard $f(T,B)$ case, we consider the three cases below which allow for progressively better cosmographic constraints on the parameters.

\begin{itemize}
    \item Case (a). Fixed $q_0$. Using the expression in Eq.~(\ref{eq:qfixed}) and the values $\tilde{t}_0 =0$ and $\tilde{b}_0= 1$ (which relates to a pure power-law in the boundary term theory), we obtain $j_0=2.630^{+0.225}_{-0.229}$ and $s_0=7.403^{+0.286}_{-0.283}$, using the full Pantheon + GRB + CC using a Markov chain Monte Carlo (MCMC) analysis.
    For this case we obtain a Universe that is in decelerating phase. This sets the first constraint on the form of $f(T,B)$, where the $q_0$ cannot be fixed if we require to follow the correct cosmic acceleration. See Fig.~(\ref{fig:qcosmo1}) for the posterior plot. Notice that once we add GRB we obtain the confidence level (C.L) overlaps on the Pantheon solely.
    \begin{figure}[ht]
    \centering
    \includegraphics[width=0.4\textwidth]{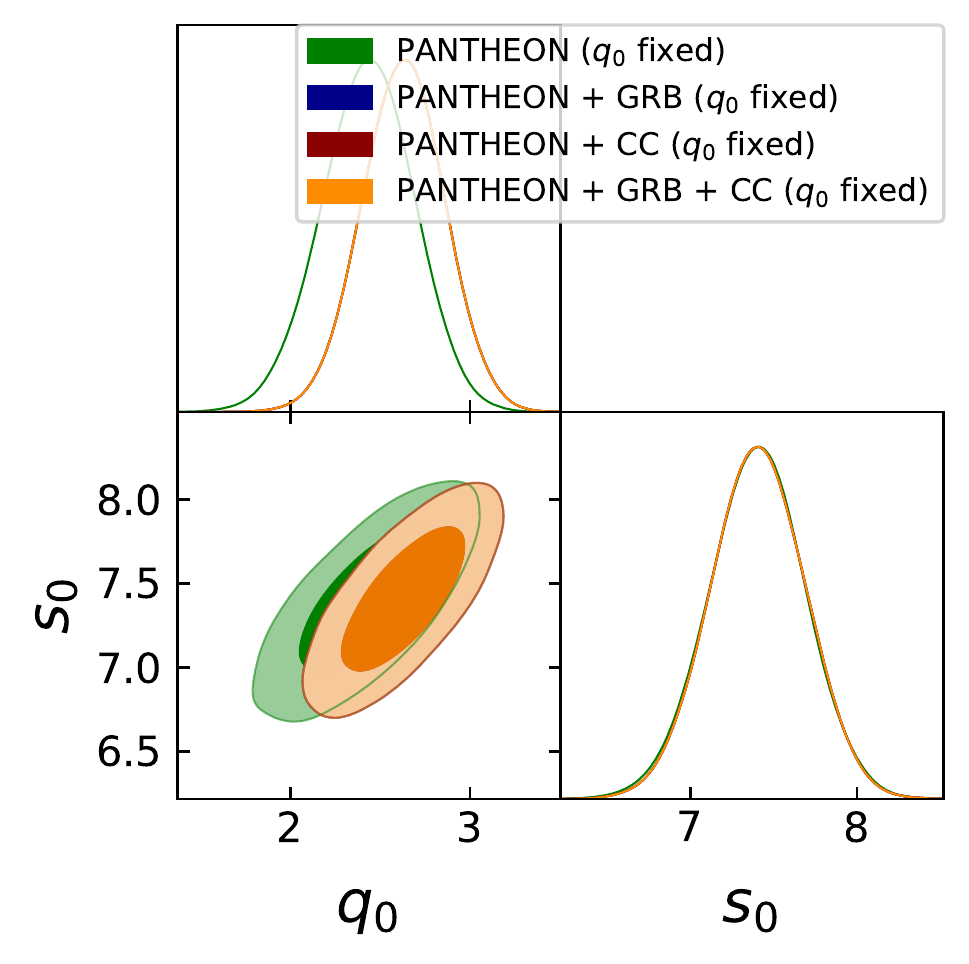}
    \caption{Case (a) using Pantheon compilation (green C.L.), Pantheon+GRB (blue C.L.), Pantheon+CC (red C.L.) and Pantheon+GRB+CC (orange C.L.).\label{fig:qcosmo1}}
\end{figure}
    \item Case (b). Fixed $j_0$. Using the expression for $j$ (\ref{eq:cosmo-fixed}) and the values $\tilde{t}_0 =0$ and $\tilde{b}_0= 1$, we obtain $q_0=-0.799^{+0.024}_{-0.025}$, $s_0=-3.409^{+0.287}_{-0.327}$, and $l_0=-8.046^{+1.973}_{-1.346}$ using the full sample Pantheon + GRB+ CC with an MCMC analysis.
    For this case we obtain a Universe that is in fact accelerating as it is expected. See Fig.~\ref{fig:qcosmo2} for details of the posterior. Here, we notice that the values of the higher derivative cosmographic parameters starts to become reasonable. 
     \begin{figure}[ht]
    \centering
    \includegraphics[width=0.5\textwidth]{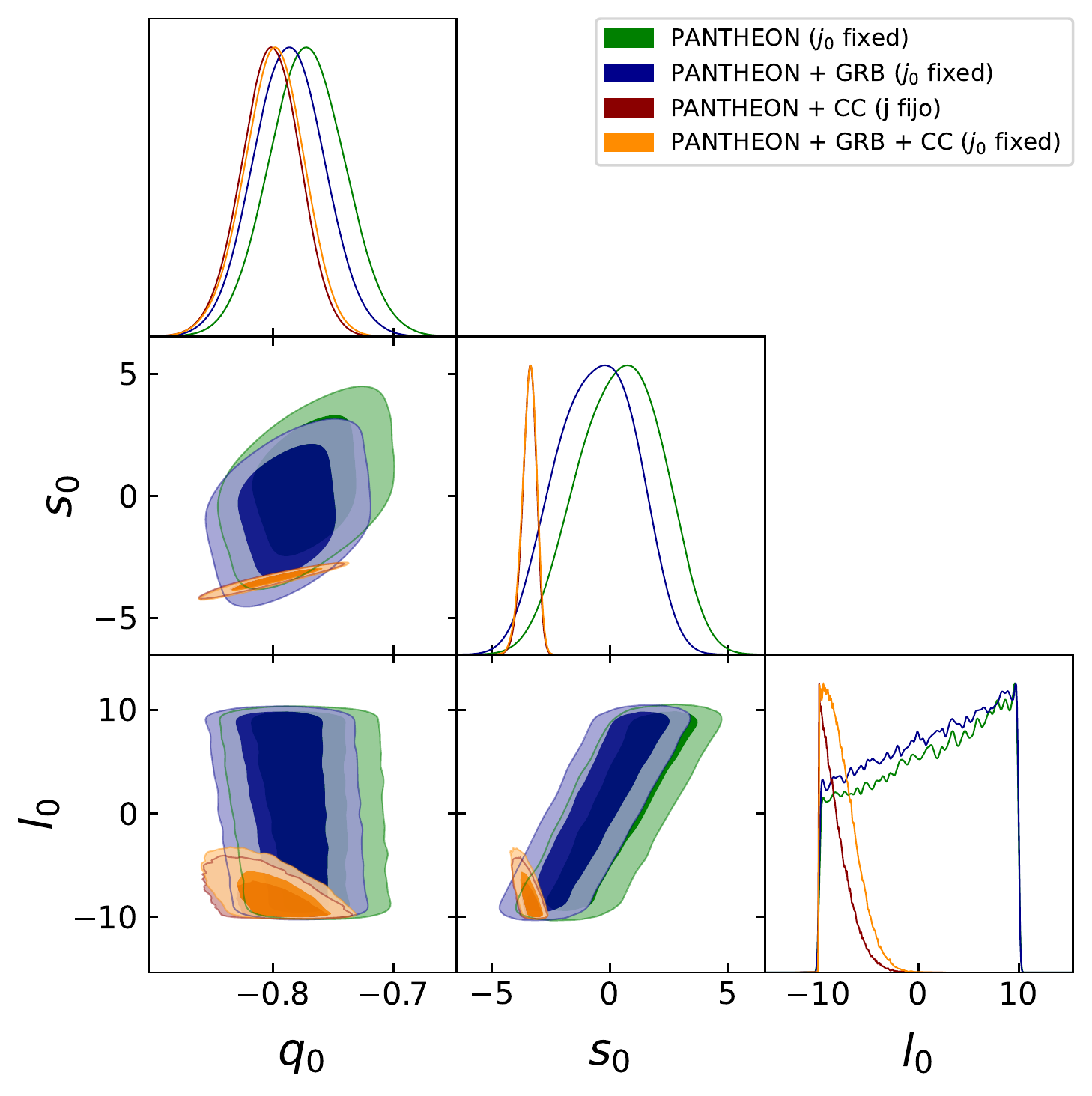}
    \caption{Case (b) using Pantheon compilation (green C.L.), Pantheon+GRB (blue C.L.), Pantheon+CC (red C.L.) and Pantheon+GRB+CC (orange C.L.).\label{fig:qcosmo2}}
    \end{figure}
    \item Case (c). Fixed $s_0$. Using the expression for s (\ref{eq:cosmo-fixed}) and the values $\tilde{t}_0 =0$ and $\tilde{b}_0= 1$, we obtain $q_0=-0.630\pm 0.018$, $j_0=1.828^{+0.115}_{-0.114}$, $l=-9.744^{+0.421}_{-0.193}$ and $m_0=9.235^{+0.571}_{-1.246}$ using the full sample Pantheon + GRB+ CC. Notice that here the cosmography can go beyond $m_0$.
    For this case we obtain a Universe that is in fact accelerating as it is expected and the addition of GRB explore the necessity of a high cosmography. See Fig.~(\ref{fig:qcosmo3}).
     \begin{figure}[ht]
    \centering
     \includegraphics[width=0.5\textwidth]{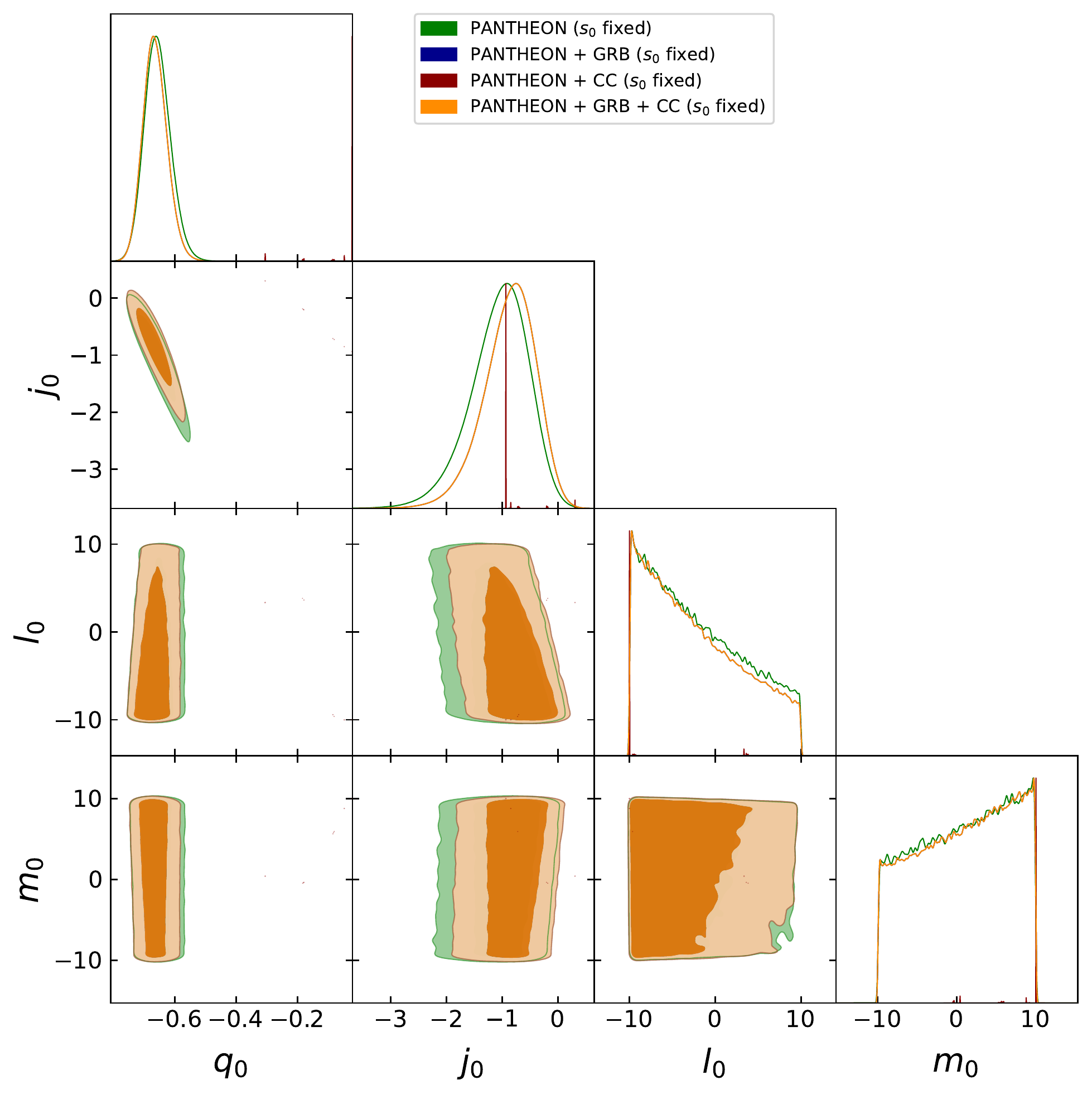}
    \caption{Case (c) using Pantheon compilation (green C.L.), Pantheon+GRB (blue C.L.), Pantheon+CC (red C.L.) and Pantheon+GRB+CC (orange C.L.).\label{fig:qcosmo3}}
    \end{figure}
\end{itemize}

\renewcommand{\arraystretch}{1.9}
\begin{table}[h]
\begin{center}
\caption{\label{cosmo_fgen}
Best cosmographic fits for the standard case using a $f(T,B)$ power law model. In first columns are denoted the three cases considered. We divided the analysis using: Pantheon SN + CC (second block column) and Pantheon SN + CC + GRB (third block column). The word \textit{fixed} indicates a flat $\Lambda$CDM prior imposed on each case.  Results are discussed in the text.
}
 \resizebox{\textwidth}{!}{  
\begin{tabular}{|c|c|c|c|c|c||c|c|c|c|c|}
\hline
\multirow{3}{*}{Cases} & \multicolumn{5}{c||}{Pantheon SN +CC} & 
    \multicolumn{5}{c|}{Pantheon SN + CC + GRB} \\
\cline{2-11}
& $q_0$ & $j_0$ & $s_0$  & $l_0$ & $m_0$ 
& $q_0$ & $j_0$ & $s_0$  & $l_0$ & $m_0$ 
\\
\hline
 {(a)}& Fixed & $2.630^{+0.225}_{-0.229}$  & $7.403^{+0.286}_{-0.283}$ &
 - & - &
Fixed & $2.710^{+0.205}_{-0.219}$  & $7.502^{+0.200}_{-0.189}$ &
-  &- \\
{(b)} & $-0.802^{+0.024}_{-0.225}$ & Fixed  & $-3.411^{+0.273}_{-0.307}$ &
 $-8.582^{+1.754}_{-1.023}$ & -&
$-0.799^{+0.224}_{-0.225}$  & Fixed  & $-3.409^{+0.115}_{-0.114}$ &
$-8.046^{+1.973}_{-1.346}$ &- \\
{(c)} & $-0.666^{0.040}_{-0.037}$ & $-0.836^{+0.419}_{-0.499}$  & Fixed &
 $-3.266^{+7.474}_{-4.878}$ 
   & $1.209^{+6.170}_{-7.295}$ 
& $-0.630\pm 0.018$ & $1.828^{+0.225}_{-0.229}$  & Fixed &
 $-9.744^{+0.421}_{-0.193}$ 
    & $9.235^{+0.571}_{-1.246}$
 \\
\hline
\end{tabular}\label{tab:results}
}
\end{center} 
\end{table}

In Table \ref{cosmo_fgen}, we show the results of the MCMC analyses using both SN+CC and SN+CC+GRB data set compilations together with the three cases considered above for the cosmographic parameters. What we find is that due to the inter-relatedness of the cosmographic parameters, some high order derivative terms cannot be constrained such as $l_0$ which id due to the high degree of interdependence of the parameters. In this context, it follows suit that we can only consider a certain amount of parameters depending on the amount of data available. Another important point concerns which parameters can be left fixed and which need to be constrained. In these analysis, we find that it is best to fix the highest derivative term and constrain the lower parameters, which leads to the best fit cosmographic parameters.

\section{Conclusions and discussion}
\label{sec:conclusions}

Cosmography was used in this work to reconstruct elements of the $f(T,B)$ Lagrangian in the context of a power law model. The cosmographic parameters (\ref{eq:Hparameter},\ref{eq:nparameter}) provide an interesting avenue to parameterizing the evolution of the Universe using progressively higher order derivative terms in the scale factor. Since they build on each other, these parameters can be related to each other which causes complicated coupling problems in many cases. For this reason, very high order derivative terms need enormous amounts of data to be constrained to any degree using observational data. On the other hand, these state parameters provide an efficient way to connect expansion terms with derivatives in a cosmological model.

In the context of TG, the cosmographic parameters can readily be injected into the scalar contributors of the theory \eqref{eq:TyB}. As expected progressively higher time derivatives of these scalar are naturally related to higher time derivatives of the Hubble parameter which is a crucial link with observations. Thus, by using the Friedmann equations, the cosmographic parameters can then be related to the form of $f(T,B)$ through the use of the Friedmann equations which is the point at which this becomes dependent on the particular model under consideration (\ref{eq:q_interms}--\ref{eq:l_interms}). The other key property to acknowledge at this point is that it not possible to isolate any of these cosmographic parameters for most model parameters values except for very specially chosen ones as explored in Case A in Eq.~\eqref{eq:dece_par_sc}. To this end, we use the luminosity distance calculation to strengthen the link to observations using the SN data set since it can be related to the distance modulus parameter, which is the true parameter of the Pantheon data set.

In Sec.~\ref{sec:data}, we consider the CC, SN and the new GRB data sets. GRB data is very interesting because it may shed new light on fundamental theory using new physical phenomena. To probe these potential differences we consider three settings of cosmographic parameters in Figs.~(\ref{fig:qcosmo1},\ref{fig:qcosmo2},\ref{fig:qcosmo3}) which respectively constraint the cosmographic parameters for fixed $q_0$, $j_0$ and $l_0$. While it is not possible in any of these analyses to completely constrain the value of high derivative terms such as the lerk $l$. Saying that, as more parameters are not fixed to a value, their values tend to the literature values such as in the case of the deceleration parameter which readily turns from a late time decelerating solution in Fig.~(\ref{fig:qcosmo1}) to an accelerating one in Figs.~(\ref{fig:qcosmo2},\ref{fig:qcosmo3}). The complete set of results are collected in Table.~\ref{tab:results} where the impact of the GRB data on the various cosmographic parameter is shown more clearly such as its preference for a weaker acceleration in the late Universe.

It would be interesting to consider other viable models in the $f(T,B)$ gravity extension to $\Lambda$CDM such as those models as proposed in Ref.~\cite{Escamilla-Rivera:2019ulu}. However, it would make the analysis much more involved. It would also be advantageous to consider other approaches to constraining the evolution of the cosmographic parameters such as the non-parametric ones proposed in Refs.~\cite{Escamilla-Rivera:2021rbe,Bernardo:2021mfs,Briffa:2020qli}.


\begin{acknowledgments}
CE-R acknowledges the Royal Astronomical Society as FRAS 10147. CE-R and GRF are supported by DGAPA-PAPIIT-UNAM Project IA100220. GRF acknowledges financial support from CONACyT postgraduate grants program. The simulations were performed in Centro de c\'omputo Tochtli-ICN-UNAM as part of CosmoNag. JLS would also like to acknowledge funding support from Cosmology@MALTA which is supported by the University of Malta. The authors would like to acknowledge networking support by the COST Action CA18108. 
\end{acknowledgments}

\appendix

\section{Derivations for the power law model cosmography} 
\label{app:calculations}

As we mentioned, consider a simple model as a power law model $f(T,B)=\tilde{t}_0T^m + \tilde{b}_0B^k$, where mixed partial derivatives vanish and it can be seen that its derivatives are given by the following expressions: 
\begin{align}
f_T &= m \tilde{t}_0 T^{m-1} \\
f_B &= k\tilde{b}_0 B^{k-1} \\
\dot{f}_T &= m(m-1) \tilde{t}_0 \dot{T} T^{m-2} \\
\dot{f}_B &= k(k-1) \tilde{b}_0 \dot{B} B^{k-2} \\
\ddot{f}_T &=  m(m-1)(m-2)\tilde{t}_0 \dot{T}^2 T^{m-3}+m(m-1) \tilde{t}_0 \ddot{T} T^{m-2}\\
\ddot{f}_B &= k(k-1)(k-2)\tilde{b}_0 \dot{B}^2 B^{k-3}+k(k-1)\tilde{b}_0 \ddot{B}B^{k-2}\\
\dddot{f}_T &=m(m-1)(m-2)(m-3)  \tilde{t}_0 \dot{T}^3 T^{m-4}+3 m(m-1) (m-2) \tilde{t}_0 \dot{T} \ddot{T} T^{m-3}\\&+m(m-1) \tilde{t}_0 \dddot{T} T^{m-2} \notag \\
\dddot{f}_B &=k(k-1)(k-2)(k-3)\tilde{b}_0 \dot{B}^3 B^{k-4}+3k(k-1) (k-2) \tilde{b}_0 \dot{B} \ddot{B}B^{k-3}+k(k-1) \tilde{b}_0 \dddot{B} B^{k-2} \\
f^{\text{(iv)}}_T &=m(m-1)(m-2)(m-3)(m-4) \tilde{t}_0 \dot{T}^4 T^{m-5}+6m(m-1) (m-2)(m-3)\tilde{t}_0\dot{T}^2 \ddot{T} T^{m-4}\\&+3m(m-1) (m-2) \tilde{t}_0 \ddot{T}^2 T^{m-3}+4 m(m-1) (m-2) \tilde{t}_0 \dot{T}
   \dddot{T} T^{m-3}+m(m-1)\tilde{t}_0T^{\text{(iv)}} T^{m-2} \notag \\
f^{\text{(iv)}}_B&=k(k-1)(k-2)(k-3)(k-4)\tilde{b}_0 \dot{B}^4 B^{k-5}+6k(k-1)(k-2)(k-3)\tilde{b}_0 \dot{B}^2 \ddot{B}B^{k-4}\\ &+3 k(k-1) (k-2) \tilde{b}_0 \ddot{B}^2 B^{k-3}+4k(k-1) (k-2) \tilde{b}_0 \dot{B}
   \dddot{B}B^{k-3}+k(k-1)\tilde{b}_0B^{\text{(iv)}} B^{k-2} \notag \\
f^{\text{(v)}}_B&= k(k-1) (k-2) (k-3) (k-4) (k-5)\tilde{b}_0 \dot{B}^5 B^{k-6}+10k(k-1) (k-2) (k-3) (k-4) \tilde{b}_0 \dot{B}^3 \ddot{B} B^{k-5}\\ \notag &+15k(k-1) (k-2) (k-3) \tilde{b}_0 \ddot{B} \dddot{B} 
   B^{k-3}+5k(k-1)(k-2) \tilde{b}_0 \dot{B}B^{\text{(iv)}} B^{k-3}+k(k-1)\tilde{b}_0 B^{\text{(v)}} B^{k-2}.
\end{align}
These derivatives can be written in terms of the cosmographic parameters \eqref{eq:Hparameter}-\eqref{eq:nparameter} using the expressions in \eqref{eq:TyB} as:
{\footnotesize
\begin{align}
f_T &= 6^{m-1} m \tilde{t}_0 \left(H^2\right)^{m-1} \label{eq:f_T}\\
f_B &= \tilde{b}_0 k \left(18 H^2-6 H^2 (q+1)\right)^{k-1} \\
\dot{f}_T &= H^3 \left(-2^m\right) 3^{m-1} (m-1) m (q+1) \tilde{t}_0 \left(H^2\right)^{m-2} \\
\dot{f}_B &= \tilde{b}_0 (k-1) k \left(6 H^3 (j+3 q+2)-36 H^3 (q+1)\right) \left(18 H^2-6 H^2 (q+1)\right)^{k-2}\\
\ddot{f}_T &= H^6 2^{m+1} 3^{m-1} (m-2) (m-1) m (q+1)^2 \tilde{t}_0 \left(H^2\right)^{m-3} \\ \notag &+H^4 2^m 3^{m-1} (m-1) m \tilde{t}_0 \left(H^2\right)^{m-2} (j+q (q+5)+3) \\
\ddot{f}_B &= \tilde{b}_0 (k-1) k \left(18 H^2-6 H^2 (q+1)\right)^{k-2} \left(6 H^4 (-4 j-3 q (q+4)+s-6)+36 H^4 (j+q (q+5)+3)\right)\\&+\tilde{b}_0 (k-2) (k-1) k \left(6 H^3 (j+3 q+2)-36 H^3 (q+1)\right)^2
   \left(18 H^2-6 H^2 (q+1)\right)^{k-3} \notag \\
\dddot{f}_T &= H^9 \left(-2^{m+2}\right) 3^{m-1} (m-3) (m-2) (m-1) m (q+1)^3 \tilde{t}_0 \left(H^2\right)^{m-4}\\&-H^7 2^{m+1} 3^m (m-2) (m-1) m (q+1) \tilde{t}_0 \left(H^2\right)^{m-3} (j+q (q+5)+3)\notag \\&+H^5 2^m
   3^{m-1} (m-1) m \tilde{t}_0 \left(H^2\right)^{m-2} (-j (3 q+7)-3 q (4 q+9)+s-12) \notag \\
\dddot{f}_B &= \tilde{b}_0 (k-1) k \left(18 H^2-6 H^2 (q+1)\right)^{k-2} \left(6 H^5 (10 j (q+2)+l+30 q (q+2)-5 s+24)\right. \\ \notag &\left.+36 H^5 (-j (3 q+7)-3 q (4 q+9)+s-12)\right)+\tilde{b}_0 (k-3) (k-2) (k-1) k \left(6 H^3 (j+3
   q+2) \right. \\ \notag &\left.-36 H^3 (q+1)\right)^3 \left(18 H^2-6 H^2 (q+1)\right)^{k-4}+3 \tilde{b}_0 (k-2) (k-1) k \left(6 H^3 (j+3 q+2)-36 H^3 (q+1)\right) \left(18 H^2\right. \\ \notag &\left.-6 H^2 (q+1)\right)^{k-3} \left(6 H^4
   (-4 j-3 q (q+4)+s-6)+36 H^4 (j+q (q+5)+3)\right) \\
f^{\text{(iv)}}_T &= H^{12} 2^{m+3} 3^{m-1} (m-4) (m-3) (m-2) (m-1) m (q+1)^4 \tilde{t}_0 \left(H^2\right)^{m-5}\\&+H^{10} 2^{m+3} 3^m (m-3) (m-2) (m-1) m (q+1)^2 \tilde{t}_0 \left(H^2\right)^{m-4} (j+q (q+5)+3)\notag \\ &-H^8
   2^{m+3} 3^{m-1} (m-2) (m-1) m (q+1) \tilde{t}_0 \left(H^2\right)^{m-3} (-j (3 q+7)-3 q (4 q+9)+s-12)\notag \\&+H^8 2^{m+1} 3^m (m-2) (m-1) m \tilde{t}_0 \left(H^2\right)^{m-3} (j+q (q+5)+3)^2\notag \\ &+H^6 2^m
   3^{m-1} (m-1) m \tilde{t}_0 \left(H^2\right)^{m-2} \left(j (3 j+44 q+48)+l+3 q \left(4 q^2+39 q+56\right)-(4 q+9) s+60\right) \notag\\
f^{\text{(iv)}}_B &= \tilde{b}_0 (k-1) k \left(18 H^2-6 H^2 (q+1)\right)^{k-2} \left(6 H^6 \left(-10 j^2-120 j (q+1)\right.\right. \\ \notag & \left. \left.-3 \left(2 l+5 \left(2 q^3+18 q^2-q s+24 q-2 s+8\right)\right)+m\right)+36 H^6 \left(j (3
   j+44 q+48)+l\right.\right. \\ \notag &\left.\left.+3 q \left(4 q^2+39 q+56\right)-(4 q+9) s+60\right)\right)+3 \tilde{b}_0 (k-2) (k-1) k \left(18 H^2-6 H^2 (q+1)\right)^{k-3} \left(6 H^4 (-4 j\right. \\ \notag & \left.-3 q (q+4)+s-6)+36 H^4 (j+q
   (q+5)+3)\right)^2+\tilde{b}_0 (k-4) (k-3) (k-2) (k-1) k \left(6 H^3 (j+3 q+2)\right. \\ \notag &\left.-36 H^3 (q+1)\right)^4 \left(18 H^2-6 H^2 (q+1)\right)^{k-5}+4 \tilde{b}_0 (k-2) (k-1) k \left(6 H^3 (j+3 q+2)-36
   H^3 (q+1)\right) \left(18 H^2 \right. \\ &\left.-6 H^2 (q+1)\right)^{k-3} \left(6 H^5 (10 j (q+2)+l+30 q (q+2)-5 s+24)+36 H^5 (-j (3 q+7)-3 q (4 q+9)+s-12)\right)\notag \\ \notag &+6 \tilde{b}_0 (k-3) (k-2) (k-1) k
   \left(6 H^3 (j+3 q+2)-36 H^3 (q+1)\right)^2 \left(18 H^2-6 H^2 (q+1)\right)^{k-4} \left(6 H^4 (-4 j \right. \\ \notag &\left. -3 q (q+4)+s-6)+36 H^4 (j+q (q+5)+3)\right)\\
f^{\text{(v)}}_B &= \tilde{b}_0 (k-1) k \left(18 H^2-6 H^2 (q+1)\right)^{k-2} \left(6 H^7 \left(140 j^2+35 j (6 q (q+6)-s+24)+21 l (q+2) \right.\right. \label{eq:f_biv}\\ \notag &\left. \left.-7m+n -210 (q+1) s+90 \left(7 q (q+2)^2+8\right)\right)-36 H^7 \left(50
   j^2+10 j (q (8 q+51)-s+36)+l (5 q+11)-m\right.\right. \\ \notag &\left. \left.-5 (14 q+15) s+30 q (q+2) (9 q+20)+360\right)\right)+\tilde{b}_0 (k-5) (k-4) (k-3) (k-2) (k-1) k \left(6 H^3 (j+3 q+2)\right. \\ \notag &\left.-36 H^3 (q+1)\right)^5
   \left(18 H^2-6 H^2 (q+1)\right)^{k-6}+5 \tilde{b}_0 (k-2) (k-1) k \left(6 H^3 (j+3 q+2)-36 H^3 (q+1)\right) \left(18 H^2 \right. \\ \notag &\left. -6 H^2 (q+1)\right)^{k-3} \left(6 H^6 \left(-10 j^2-120 j
   (q+1)-3 \left(2 l+5 \left(2 q^3+18 q^2-q s+24 q-2 s+8\right)\right)+m\right)\right. \\ \notag &\left. +36 H^6 \left(j (3 j+44 q+48)+l+3 q \left(4 q^2+39 q+56\right)-(4 q+9) s+60\right)\right)+15 \tilde{b}_0
   (k-3) (k-2) (k-1) k \left(18 H^2\right. \\ \notag &\left.-6 H^2 (q+1)\right)^{k-3} \left(6 H^4 (-4 j-3 q (q+4)+s-6)+36 H^4 (j+q (q+5)+3)\right) \left(6 H^5 (10 j (q+2)+l\right. \\ \notag &\left.+30 q (q+2)-5 s+24)+36 H^5 (-j
   (3 q+7)-3 q (4 q+9)+s-12)\right)\\ \notag &+10 \tilde{b}_0 (k-4) (k-3) (k-2) (k-1) k \left(6 H^3 (j+3 q+2)-36 H^3 (q+1)\right)^3 \left(18 H^2-6 H^2 (q+1)\right)^{k-5}\\ \notag & \left(6 H^4 (-4 j-3 q
   (q+4)+s-6)+36 H^4 (j+q (q+5)+3)\right). 
\end{align}
}


\section{Phenomenological Considerations}\label{app_pheno}
In this work we considered beforehand the model $f(T,B) = -T + \tilde{f}(T,B)$ with $\tilde{f}(T,B) = \tilde{t}_0T^k + \tilde{b}_0 B^m$. This model is too complex to be generally analysed 
in the context of cosmography, hence, we decided to work with one of the simplest form of these kind of models, the model where $\tilde{t}_0=0$, $\tilde{b}_0=1$ and $m=2$, which basically makes the functional to be $f(T,B)= -T + B^2$. Although this model can be somehow justified from a theoretical point of view (in the sense that it is a simple extension of TEGR), it suffers from an ill phenomenology. Let us see this in detail.

The first modified Friedman equation in $f(T,B)$ with the signature $(-,+,+,+)$ is given by (see Eq. (19) in \cite{Escamilla-Rivera:2019ulu})
\begin{align}
-3H^2(3f_B + 2f_T) + 3H\dot{f}_B - 3\dot{H}f_B + \frac{1}{2}f = \kappa^2 \rho_m. \label{firstfriedman-}
\end{align}
Since the boundary term and the torsion scalar are $B = 6(3H^2 + \dot{H})$ and $T = 6H^2$ respectively, the Eq. \eqref{firstfriedman-} can be rewritten as
\begin{align}
-\frac{1}{2}Bf_B  - Tf_T + 3H\dot{f}_B + \frac{1}{2}f = \kappa^2 \rho_m. \label{firstfriedman-v2}
\end{align}
Consider the model $f(T,B) = -T + \alpha B^2$, which is exactly the same used here if $\alpha=1$. With this model, the Eq. \eqref{firstfriedman-v2} becomes
\begin{align}
-\frac{1}{2}B(2\alpha B)  - T(-1) + 3H(2\alpha \dot{B}) + \frac{1}{2}(-T + \alpha B^2) = \kappa^2 \rho_m. \label{firstfriedman-v3}
\end{align}
Using the Eqs.\eqref{eq:TyB}, the form of $T$ and $B$, the Eq. \eqref{firstfriedman-v3} evaluated at  $z=0$ is
\begin{align}
6 \alpha  H_0 \left(6 H_0^3 (j_0+3 q_0+2)-36 H_0^3 (q_0+1)\right)-\alpha 
   \left(18 H_0^2-6 H_0^2 (q_0+1)\right)^2+\frac{1}{2} \left(\alpha 
   \left(18 H_0^2-6 H_0^2 (q_0+1)\right)^2\right. \\ \left. -6 H_0^2\right)+6 H_0^2= 3H_0^2 \Omega_{m0}, \notag
\end{align}
which can be solved in terms of $\Omega_{m0}$ as
\begin{align}
\Omega _{m0}= -72 \alpha  H_0^2+12 \alpha  H_0^2 j_0-6 \alpha  H_0^2
   q_0^2-12 \alpha  H_0^2 q_0+1, \label{omega}
\end{align}
or from the amplitude coefficient equation $\alpha$ as
\begin{align}
\alpha = -\frac{1-\Omega _{m0}}{6 H_0^2 \left(2 j_0-q_0^2-2 q_0-12\right)}. \label{alpha}
\end{align}
If we substitute $\alpha = 1$ which corresponds to our simplified model and the values of the cosmographic parameters corresponding to $\Lambda$CDM ($q_0 = -0.55$, $H_0 = 67.3$ and $j_0=1$), the density parameter is
\begin{align}
\Omega_{m0} \approx 2.5 \times 10^4.
\end{align}
On the other hand, if we substitute $q_0 = -0.55$, $H_0 = 67.3$, $j_0=1$ and $\Omega_{m0}=0.315$ into \eqref{alpha}, the amplitude coefficient is
\begin{align}
\alpha \approx 2.7391 \times 10^{-6}. 
\end{align} 
Hence, even though the model can recover a reasonable set of values for the cosmographic parameters, it is not completely \textit{phenomenological} correct since it leads to a wrong value of the critical matter density parameter. On the other hand, this model does have some motivation from observational constraint analysis  \cite{Escamilla-Rivera:2019ulu}.

\bibliographystyle{utphys}
\bibliography{refs}

\end{document}